\documentclass[conference]{IEEEtran}
\usepackage{epsfig,graphics,graphicx,subfigure,amssymb,amstext,amsmath,algorithm,
algorithmic,wrapfig,multirow}
\usepackage{cite}
\usepackage{url}
\usepackage{times}
\usepackage{float}
\usepackage{placeins}
\usepackage{textcomp}
\usepackage[font=small,bf]{caption}
\usepackage{hyperref}
\usepackage{flushend}
\usepackage{xcolor}
\usepackage{tabularx}
\usepackage{bm}

\def\beq{\begin{equation}}
\def\eeq{\end{equation}}
\def\beqa{\begin{eqnarray}}
\def\eeqa{\end{eqnarray}}
\def\beqan{\begin{eqnarray*}}
\def\eeqan{\end{eqnarray*}}

\def\C{{\mathbb{C}}}
\def\argmin{\mathop{\mathrm{arg\,min}}}

\def\diag{\mathop{\mathrm{diag}}}

\def\Tr{\mathop{\mathrm{Tr}}}
\def\x{\times}

\newtheorem{proposition}{Proposition}

\newtheorem{lemma}{Lemma}

\def\Exp{\mathbb{E}}

\def\tm1{t\! - \! 1}
\def\tp1{t\! + \! 1}
\def\km1{k\! - \! 1}
\def\kp1{k\! + \! 1}

\def\zerobf{\mathbf{0}}

\def\dbf{\mathbf{d}}

\def\hbf{\mathbf{h}}
\def\pbf{\mathbf{p}}

\def\qbf{\mathbf{q}}
\def\qbfhat{\widehat{\mathbf{q}}}
\def\rbf{\mathbf{r}}

\def\sbf{\mathbf{s}}

\def\ubf{\mathbf{u}}

\def\vbf{\mathbf{v}}

\def\wbf{\mathbf{w}}
\def\wbfhat{\widehat{\mathbf{w}}}

\def\ybf{\mathbf{y}}
\def\zbf{\mathbf{z}}

\def\Abf{\mathbf{A}}
\def\Bbf{\mathbf{B}}

\def\Dbf{\mathbf{D}}

\def\Gbf{\mathbf{G}}

\def\Ibf{\mathbf{I}}
\def\fbar{\overline{f}}
\def\Jbar{\overline{J}}

\def\Pbf{\mathbf{P}}

\def\Qbf{\mathbf{Q}}
\def\Qbfhat{\widehat{\mathbf{Q}}}
\def\Rbf{\mathbf{R}}

\def\Sbf{\mathbf{S}}

\def\Ubf{\mathbf{U}}

\def\alphabar{\overline{\alpha}}

\newif\ifconf
\conffalse

\IEEEoverridecommandlockouts

\renewcommand{\footnoterule}{%
  \kern -3pt
  \hrule width \columnwidth height 0.5pt
  \kern 3pt
}
\begin{document}

\title{Low-Rank Spatial Channel Estimation\\
for Millimeter Wave Cellular Systems}

\ifconf
\else
    \author{
        Parisa A. Eliasi,~\IEEEmembership{Student Member,~IEEE},
        Sundeep Rangan,~\IEEEmembership{Senior Member,~IEEE},\\
        Theodore S. Rappaport,~\IEEEmembership{Fellow,~IEEE}
        \thanks{This material is based upon work supported by the National Science
        Foundation under Grants No. 1116589 and 1237821 as well as generous support
        from Samsung, Nokia Siemens Networks, Intel, Qualcomm and InterDigital Communications.}
        \thanks{
            The authors are with NYU WIRELESS, New York University Polytechnic
            School of Engineering, Brooklyn, NY 11201 USA
            (e-mail: pa854@nyu.edu;
            srangan@nyu.edu; tsr@nyu.edu).}
    }
\fi

\maketitle

\begin{abstract}
The tremendous bandwidth available in the
millimeter wave (mmW) frequencies between
30 and 300~GHz have made these bands an attractive candidate
for next-generation cellular systems.
However, reliable communication at these frequencies
depends extensively
on beamforming with very high-dimensional antenna arrays.
Estimating the channel sufficiently accurately to perform
beamforming can thus be challenging both
due to low coherence time and large number of antennas.
Also, the measurements used for channel estimation
may need to be made with analog beamforming
where the receiver can ``look" in only direction at a time.
This work presents a novel method for estimation of
the receive-side spatial covariance matrix of a channel
from a sequence
of power measurements made at different angular directions.
The method reduces the spatial covariance estimation to a
matrix completion optimization problem.  To reduce the number of
measurements, the optimization can incorporate the
low-rank constraints in the channels that are typical
in the mmW setting.
The optimization is convex and fast, iterative methods are presented
to solving the problem.
Simulations  are presented for both single and multi-path channels
using channel models derived from real measurements in New York City
at 28~GHz..
\end{abstract}

\section{Introduction}

Meeting Tthe tremendous growth in demand
for cellular data~\cite{CiscoVNI:latest}
will require
 new technologies that can provide orders of magnitude increases in wide-area
 wireless capacity.
With the severe shortage of spectrum in traditional UHF and microwave
bands below 3~GHz, there has been considerable interest
in so-called millimeter wave (mmW) frequencies
between 30 and 300 GHz where vast amounts of essentially virgin
spectrum are still widely available
~\cite{KhanPi:11-CommMag,PietBRPC:12,rappaportmillimeter,RanRapE:14,BocHLMP:14,Rappaport2014-mmwbook}.

However, a significant challenge for using mmW for wide-area,
cellular-type coverage is range.
Due to Friis' Law ~\cite{Rappaport:02}, the high frequencies
of mmW signals result in large isotropic path loss.
Fortunately, the very small wavelengths of mmW signals combined
with advances in low-power CMOS RF circuits enable large numbers
($\geq$ 32 elements) of miniaturized antennas to be placed in small
dimensions thereby providing high beamforming gains that can
 theoretically more than compensate for the increase in isotropic path loss
\cite{AkdenizCapacity:14}.

\begin{figure}
\centering {\includegraphics[trim=5.5cm 4.5cm 3cm 4.5cm ,clip=true, width=1\linewidth]{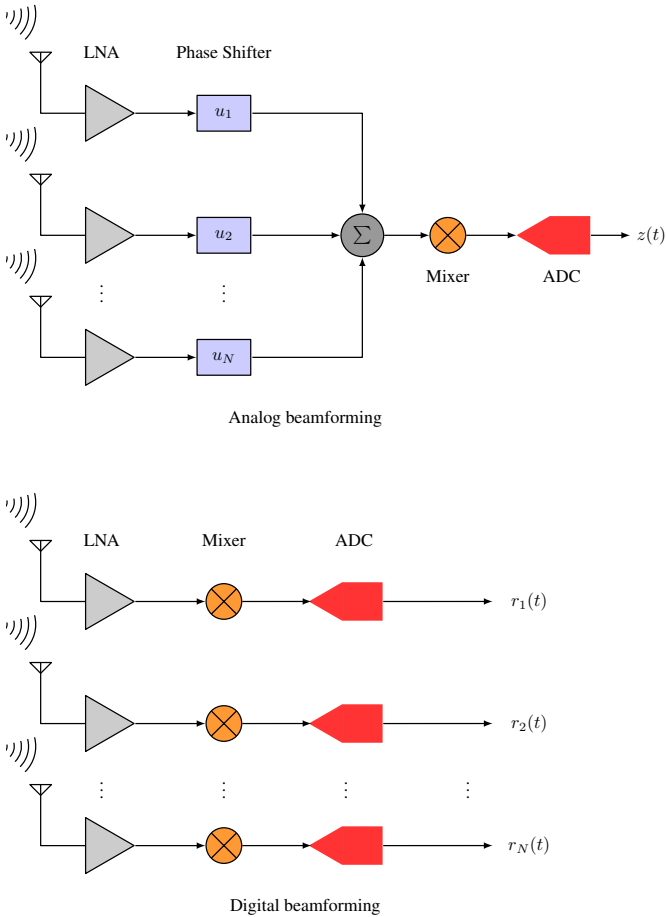}}
\caption{\textbf{Analog vs.\ digital beamforming.}
\textbf{Bottom panel}:
Front-ends at conventional frequencies typically digitize the signals from
each antenna separately. This fully digital architecture offers the greatest
flexibility.  However, power consumption may be prohibitive in the mmW range
when the bandwidth and number of antennas is large.  \textbf{Top panel}:
To reduce power consumption, mmW front-end receivers
may need to perform beamforming in analog via phase shifters.
A consequence of this architecture for channel estimation is that each measurement
provides information in only one direction at a time.
}
\label{fig:analogBF}
\end{figure}

However, spatial channel estimation needed to support
beamforming presents several challenges in the mmW range:
\begin{itemize}
\item \emph{High-dimensional arrays:}
Since current mobile devices typically have one to four antennas,
the array sizes in the mmW range -- which may be 16 or 32 elements
even at the mobile -- will represent an significant
increase in the dimension of the antenna processing.
In particular, a much larger number of parameters
will need be tracked at the receiver for channel estimation.
A system with $N_{rx}$ receive antennas must
estimate $N_{rx}$ channels per transmit
stream for instantaneous beamforming
and $N_{rx}^2$ parameters for the
receive-side spatial covariance matrix used in long-term beamforming.

\item \emph{Rapid channel variations:}
The high frequencies of the mmW bands implies that the
coherence time of the channel may be very small,
meaning that each of the channels to be tracked can be
varying rapidly.
Channel tracking for small-scale fading can be avoided
by long-term beamforming~\cite{Lozano:07}, and simulations
based on experimental
measurements in \cite{AkdenizCapacity:14}
suggest that the long-term
beamforming introduces only a 1 to 2 dB loss in the mmW range.
However, since mmW signals are extremely susceptible to
blocking \cite{Rappaport2014-mmwbook},
even the large-scale channel characteristics may change
rapidly.
For example, a change in the orientation of the mobile
device, movement of a hand holding the device
or appearance of a wall would all change the channel
significantly.
Thus, channel statistics must be estimated with a limited
number of measurements.

\item \emph{Analog beamforming}:
Due to the high bandwidths and large number
of antenna elements in the mmW range,
it may not be possible from a power consumption perspective
for the mobile receiver to obtain high rate digital samples
from all antenna elements~\cite{KhanPi:11}.
Most proposed designs perform beamforming in analog (either in RF or IF)
prior to the A/D conversion
\cite{KohReb:07,KohReb:09,GuanHaHa:04,Heath:partialBF}
-- see Fig.\ \ref{fig:analogBF}.  A key limitation for these
architectures is that they permit the
mobile to ``look" in only one or a small number of directions at a time. This feature significantly reduces the information
in each measurement, further complicating the channel estimation
process.
\end{itemize}

In this paper, we consider the problem of estimating
the long-term receive-side spatial covariance of a channel
on a high-dimensional array from a limited number of
analog measurements.
Key to our methodology is that the mmW channels
will likely have a low-rank structure relative to the
number of antenna elements.
For example, extensive measurements at 28 and 73~GHz in New York City
\cite{Rappaport:12-28G,Samimi:AoAD,rappaportmillimeter}
-- a dense, urban environment similar to likely initial deployments
for mmW systems --
have shown that the mmW channel energy is often concentrated
in a small number of relatively narrow-beam clusters.
Analysis of this data in \cite{AkdenizCapacity:14}
have revealed that the channel is often well approximated
by a rank three or four channel, typically much smaller
than the antenna dimension.
Similar findings can be found in \cite{molisch2014propagation}.
This low-rank property implies
that the spatial covariance matrix can be characterized by
a relatively small number of
parameters for the purpose of channel estimation.

Of course, the use of low-rank spatial structure is widely-used
in array processing and underlies many classic
channel estimation for wireless systems
\cite{martone1998adaptive,ottersten1996array,wang1998blind}.
The contribution in this work is to consider
the use of low-rank channel estimation
from analog measurements.  As we describe below,
each measurement from an array with analog phase shifting
provides a power measurement in a single angular direction.
We show that maximum likelihood (ML)
reconstruction of the channel covariance matrix
from a collection of such measurements made at random angles is similar to a
low-rank matrix completion problem that has been used
widely in machine learning and image processing.

There are now several algorithms to solve low-rank matrix reconstruction
--- most are either based on nuclear or trace norm regularization
\cite{wright2009robust,lin2010augmented,koltchinskii2011nuclear}
or message passing techniques \cite{keshavan2010matrix,rangan2012iterative}.
A recent work \cite{chenrobust} has also considered low-rank recovery
problem specifically for covariance matrix estimation.
In this paper, we adapt a simple iterative soft thresholding algorithm (ISTA) method
\cite{BeckTeb:09} originally used in sparse recovery problems,
but also used for matrix completion \cite{cai2010singular}.
The method here is modified to account for the non-Gaussian nature of the power
measurements in the ML objective.  It is shown that the proposed ISTA-based
algorithm converges to the global maxima of the likelihood.

Unfortunately, similar to the original work \cite{cai2010singular},
the thresholding step in each iteration of the proposed ISTA method requires an eigenvalue
decomposition of the current covariance matrix estimate.  This computation
may be preclude implementation for real-time system.  We thus propose an alternate
approximate ML estimate, where the search is performed over a appropriately
chosen finite subspace.  We show that the resulting optimization for the
approximate ML estimate is equivalent to an inference problem
for a generalized linear model (GLM) \cite{NelWed:72} with non-negative
components.
An similar ISTA method can be used to solve this GLM-type optimization,
using simple scalar thresholding avoiding all eigenvalue decompositions.

Both the exact and approximate ML algorithms are tested in both single-path
and multi-path models.  The channels for the multipath test scenarios
are from \cite{AkdenizCapacity:14} based on 28~GHz New York City data mentioned above
\cite{Rappaport:12-28G,Rappaport:28NYCPenetrationLoss,Samimi:AoAD,rappaportmillimeter}.
It is shown that the exact ML method offers excellent performance in a relatively
small number of iterations and the approximate ML method is only slightly worse.

\section{Problem Formulation}

The problem is to estimate the second-order spatial
statistics between a transmitter (TX) and receiver (RX).
We assume that the TX sends data from a single antenna, or equivalently,
from multiple TX antennas with a fixed beamforming vectors.  The RX has $N$ antennas,
and makes $L$ measurements.
In each  measurement $\ell$, $\ell=1,\ldots,L$,
the TX sends $D$ waveforms, $p_{\ell d}(t)$, $d=1,\ldots,D$,
potentially at the same time, but at different frequencies.

\begin{figure}
\centering {\includegraphics[trim=5cm 16cm 5cm 4.5cm ,clip=true, width=1\linewidth]{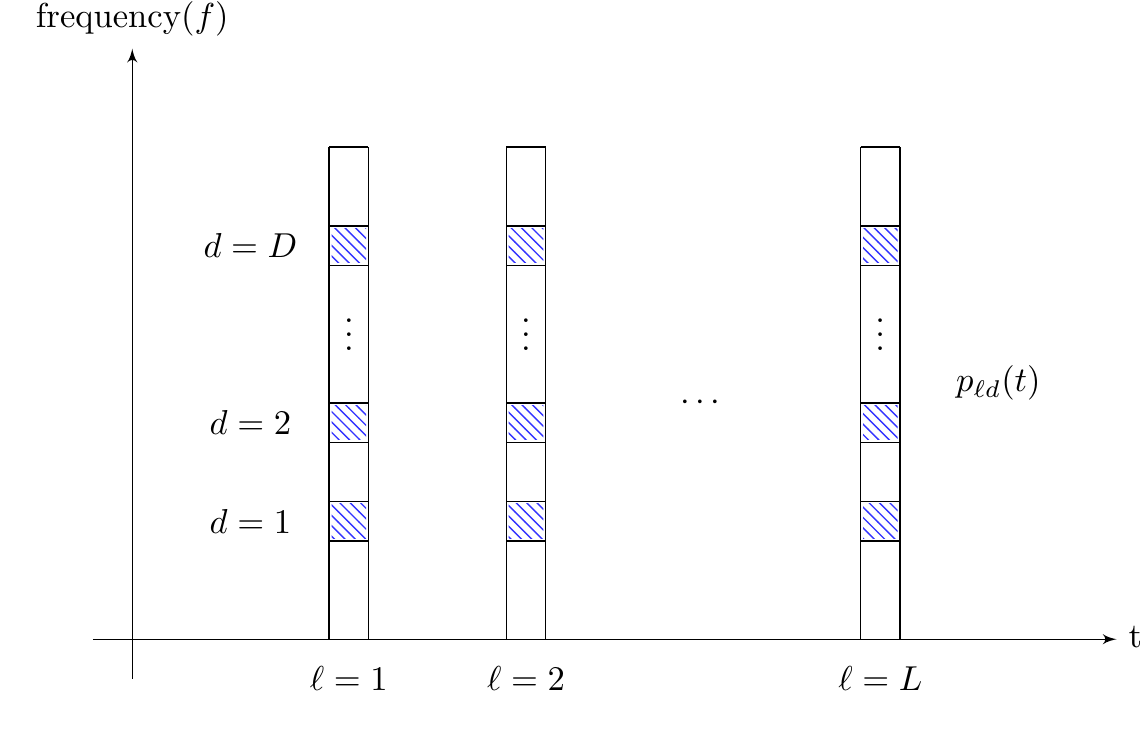}}
\caption{Model for the synchorization signals from which the
spatial channel must be estimated.  The signal is transmitted $L$ time slots,
and, for frequency diversity, the signal may be transmitted in $D$ different frequencies
in each time slot.  We will evaluate the estimation performance as a function
of $D$ and $L$.}
\label{fig:PSSsignal}
\end{figure}

An example transmission scheme is illustrated in Fig.~\ref{fig:PSSsignal}.
In this example, the transmissions are separated in time as would occur for periodic
synchronization signals such as those proposed for the Primary Synchronization Signal in
\cite{barati2014directional}.  However, the method proposed here would equally apply
to measurements from a continuous sequence of time slots such as cell
reference signals.

We assume the received complex baseband
signal across the $N$ antenna from the transmission is given by
the vector $\rbf_\ell(t) \in \C^N$, where
\beq \label{eq:rell}
    \rbf_\ell(t) = \frac{1}{\sqrt{D}}\sum_{d=1}^D \hbf_{\ell d}p_{\ell d}(t) + \vbf_\ell(t),
\eeq
where $\hbf_{\ell d}$ is the channel gain vector for the signal
$p_{\ell d}(t)$ and $\vbf_\ell(t)$ is complex AWGN with noise PSD $N_0$ Watts/Hz.
Implicitly, we assume in this model that each $p_{\ell d}(t)$ is transmitted
in a sufficiently small time and frequency region that the channel is flat across
the transmission. The factor $1/\sqrt{D}$ is used to normalized the power.

We will assume a standard \emph{correlated Rayleigh fading model} \cite{TseV:07}, where
the instantaneous channel gains $\hbf_{\ell d}$ have complex Gaussian distributions
\beq \label{eq:hgauss}
    \hbf_{\ell d} \sim {\mathcal CN}(0,\Qbf), \quad
        \Qbf = \Exp(\hbf_{\ell d}\hbf_{\ell d}^*),
\eeq
for some spatial covariance matrix $\Qbf$.
In addition, we will assume that in each measurement $\ell$,
the channel is independently faded across the different
transmissions, $\hbf_{\ell d}$, $d=1,\ldots,D$.  We thus call the parameter $D$
the \emph{diversity order}.

In this paper, we do not consider the problem of predicting
the \emph{instantaneous} channel gains $\hbf_{\ell d}$.  Instantaneous channel tracking across a large
number of antennas may be difficult in the mmW regime due to the low coherence time
and the limitation that the channel can be observed in only direction at a time
due to analog beamforming~\cite{RanRapE:14,Rappaport2014-mmwbook}.
Instead, in this paper, we thus consider only the problem of tracking the
second-order spatial statistics, namely the matrix $\Qbf$.
As described in \cite{TseV:07}, the covariance matrix
$\Qbf$ is determined by the angles of arrivals of different paths from the transmitter,
their relative average powers and
the response of the receive antenna array to the each of these paths.
Unlike the instantaneous channel gains $\hbf_{\ell d}$ which will vary due to small scale motion
(on the order of a wavelength), the long-term statistics such as $\Qbf$
depend only on the macro-layer
scattering environment and are thus a relatively constant over much longer periods of time and frequency.
In particular, in this study, we will assume that $\Qbf$ is constant over all measurements $\ell=1,\ldots,L$.

Once the spatial covariance matrix is estimated, one can perform a number of long-term
beamforming techniques~\cite{Lozano:07}.  For example,
the long-term beamforming vector that maximizes the average signal energy
can be determined  from the maximal eigenvector of $\Qbf$.
Similarly, if one estimates spatial covariance matrix $\Qbf_{sig}$
of a desired signal and the covariance matrix $\Qbf_{int}$ of the interference plus noise, the
maximal eigenvector of $\Qbf_{int}^{-1/2}\Qbf_{sig}\Qbf_{int}^{-1/2}$
is the direction that maximizes the signal-to-inference plus noise (SINR).
As mentioned in the Introduction, simulations in \cite{AkdenizCapacity:14} suggest that
the loss from optimal long-term beamforming in the mmW range relative
to instantaneous beamforming is on the order of 1~dB.

In estimating the spatial covariance matrix $\Qbf$,
our key problem assumption is that the RX does not have direct digital samples of the
components of the vector $\rbf_\ell(t)$ from the different antennas.
Instead, in each measurement $\ell$,
the RX must apply some beamforming vector $\ubf_\ell \in \C^N$ in analog and then perform
the estimation from the weighted signal
$\ubf_\ell^*\rbf_\ell(t)$.  To perform the estimation, we assume that the RX
performs a match filter with each of the signals $p_{\ell d}(t)$ to yield complex scalar outputs,
\beq \label{eq:zelld}
    z_{\ell d} = \frac{1}{\sqrt{E_s}\|\ubf_\ell\|} \int p_{\ell d}^*(t) \ubf_\ell^*\rbf_\ell(t)dt,
    \quad E_s = \int |p_{\ell d}(t)|^2dt
\eeq
where $E_s$ is the energy in the transmitted signal.
We assume that $E_s$ is the same for all $p_{\ell d}(t)$.

\section{Maximum Likelihood Estimation and Matrix Completion}\label{sec:MLE}

\subsection{Maximum Likelihood Estimation}

The problem is to estimate the spatial covariance matrix $\Qbf$
from the measurements $z_{\ell d}$.  We will assume that noise level $N_0$
is known.
We will also assume that the signals $p_{\ell d}(t)$ are orthogonal, and the channel gains
$\hbf_{\ell d}$ are independently faded across $\ell$ and $d$ and independent of the noise $\vbf(t)$.
Under these assumptions, it can be verified that the accumulated energies
\beq \label{eq:yell}
    y_\ell = \sum_{d=1}^D |z_{\ell d}|^2,
\eeq
provide a sufficient statistic for the unknown parameters
$\Qbf$ and $N_0$.  Moreover, under the independence assumptions, the
random variables $y_\ell$ will be distributed as
\[
    Y_\ell = \frac{\lambda_\ell}{2D} V_\ell,
\]
where $V_{\ell}$ is a chi-squared random variable with $2D$ degrees of freedom, and
$\lambda_\ell$ is the energy
\beq \label{eq:lamell}
    \lambda_\ell(\Qbf) = \ubf_\ell^*\left[\Qbf + \gamma^{-1} \Ibf\right]\ubf_\ell,\quad
        \gamma = \frac{E_s}{N_0}.
\eeq
See similar calculations in \cite{shiu2000fading}.
If we let $\ybf = (y_1,\ldots,y_L)$ be the vector of the received powers in the $L$ measurements.
then the negative log likelihood of $\ybf$ given $\Qbf$ is
\beq \label{eq:logp}
     -\log p(\ybf|\Qbf) = C + \sum_{\ell=1}^L \left[ D \log(\lambda_\ell(\Qbf))
        + \frac{D y_\ell}{\lambda_\ell(\Qbf)} \right],
\eeq
where $\lambda_\ell(\Qbf)$ is given in \eqref{eq:lamell}
and $C$ is some constant that does not depend on $\Qbf$ (although it may depend on $\ybf$).
Thus, we have the ML estimation of $\Qbf$ is given by
\beq \label{eq:JQopt}
    \Qbfhat = \argmin_{\Qbf} J(\Qbf) \mbox{ s.t. } \Qbf \geq \zerobf,
\eeq
where
\beq \label{eq:JQ}
     J(\Qbf) :=  \sum_{\ell=1}^L \left[ \log(\lambda_\ell(\Qbf))
        + \frac{y_\ell}{\lambda_\ell(\Qbf)} \right].
\eeq

\subsection{Connections to Matrix Completion}

An arbitrary $N \x N$ matrix $\Qbf$ has $N^2$ unknowns, and a Hermetian matrix $\Qbf=\Qbf^*$
has $N(N+1)/2$ unknowns.  Thus, one may think that one would need at least $L \geq N(N+1)/2$
measurements to fully reconstruct $\Qbf$.
However, a key property of the covariance matrix $\Qbf$ in the mmW range
is that it is typically ``almost" low-rank, meaning that the most of the energy
of the channel gains $\hbf_{\ell d}$ is concentrated in a low-dimensional
subspace.  For wireless channels, the rank of the receive-side spatial covariance matrix $\Qbf$
is determined by the number of distinct angles of arrival of paths from the
transmitter \cite{TseV:07}.
In the mmW range, analysis in \cite{AkdenizCapacity:14}
of the 28 and 73~GHz measurements in New York City in
\cite{Rappaport:12-28G,Rappaport:28NYCPenetrationLoss,Samimi:AoAD,rappaportmillimeter},
revealed that when low-power transmitters were placed in microcellular
type deployments, receivers in most
street-level locations saw only two to three dominant path clusters,
each with a relatively small angular spread.
The clustering of the paths into small, relatively narrowbeam clusters
causes the spatial covariance matrix to be low-rank.
For example, simulations in \cite{AkdenizCapacity:14}
assuming a 4x4 uniform linear array with the NYC channels showed that,
most of the energy would be likely concentrated to three to four-dimensional space
-- much lower than the 16 dimensions of the antenna array.

This low-rank property can be exploited to recover the matrix $\Qbf$ from
less than $N^2$ measurements.  To understand how this is possible,
consider the problem of
low-rank matrix completion used in machine vision and ranking systems
\cite{wright2009robust,lin2010augmented,koltchinskii2011nuclear}.
In the matrix completion problem,
one is to reconstruct a low-rank matrix $\Abf$ from a small number of entries $A_{ij}$.
If an $M \x N$ matrix $\Abf$ has rank $r$, it has only $O(r(M+N)$ degrees of freedom.
When the rank $r$ is small, this number of degrees of freedom
may be significantly less than the $MN$ parameters needed to describe a general matrix.
Low-rank matrix completion methods impose the low-rank rank constraint

In the ML estimation problem considered here, each measurement $y_\ell$ in \eqref{eq:lamell}
has an average value $\lambda_\ell$ in \eqref{eq:lamell} which is a linear
function of the unknown matrix $\Qbf$.  Hence, the ML estimation problem can be considered
a ``noisy" matrix completion problem where we attempt to attempt to reconstruct
a matrix an $N \x N$ low-rank matrix $\Qbf$ from $L$ noisy linear measurements of $\Qbf$.
The difference between the ML estimation problem considered here in contrast to
the estimation problems here

\subsection{Sparsity Regularization} \label{sec:sparseReg}

Placing a low-rank constraint on $\Qbf$ in the
optimization \eqref{eq:JQopt} will, in general, result in a non-convex
optimization.
Most matrix completion methods such as
\cite{wright2009robust,lin2010augmented,koltchinskii2011nuclear}
thus impose the low-rank constraint
indirectly by adding a regularization term of the form
$\|\Qbf\|_*$ to the objective
where $\|\Qbf\|_*$ is the so-called nuclear norm, which is the sum of the singular values
of $\Qbf$.
In our problem, the matrix $\Qbf$ is positive semi-definite,
so the nuclear norm is simply the trace: $\|\Qbf\|_* = \Tr(\Qbf)$.
We thus consider the regularized optimization
\beq \label{eq:JQoptMu}
    \Qbfhat = \argmin_{\Qbf} J_{\mu}(\Qbf) \mbox{ s.t. } \Qbf \geq \zerobf,
\eeq
where
\beq \label{eq:JQmu}
    J_{\mu}(\Qbf) := J(\Qbf) + \mu\Tr(\Qbf),
\eeq
and $\mu > 0$ is a regularization parameter.
When $\mu=0$, the objective function \eqref{eq:JQmu} agrees with the
original un-regularized ML objective \eqref{eq:JQ}.
Using $\mu > 0$ tends to enforce the requirement that $\Qbf$ is
lower rank by penalizing the eigenvalues of $\Qbf$.
In analogy with compressed sensing, the parameter $\mu > 0$ is
often considered a \emph{sparsity} regularizer since
the resulting eigenvalues of the optimal solution $\Qbfhat$
in \eqref{eq:JQoptMu} tend to have a sparse set of eigenvalues,
meaning that many will be zero
\cite{wright2009robust,lin2010augmented,koltchinskii2011nuclear}.

Interestingly, in the simulations below, we will see that $\mu > 0$
appears to not improve the performance over $\mu=0$.  This phenomena is
significantly different than the standard matrix
completion problem where using $\mu > 0$ is essential.
However, the ML objective \eqref{eq:JQopt}
already imposes a positivity constraint $\Qbf > 0$.
It is known that in related problems \cite{lee2001algorithms},
that non-negativity constraints already tends to result in sparse solutions with many
zero values -- so it is not surprising that using $\mu > 0$ does not help.

\section{Optimization Methods} \label{sec:optim}

\subsection{ISTA Algorithm}\label{sec:optimISTA}

The objective function $J(\Qbf)$ in \eqref{eq:JQ}
in the optimization \eqref{eq:JQopt},
or the sparsity-regularized version $J_{\mu}(\Qbf)$ in \eqref{eq:JQmu},
are both convex and therefore can be minimized via a number of methods.
We will first consider a simple ISTA approach \cite{BeckTeb:09}
used commonly in compressed sensing.  We will describe the ISTA algorithm
for the sparsity-regularized objective function $J_{\mu}(\Qbf)$
in \eqref{eq:JQmu}.  The algorithm for minimizing $J(\Qbf)$ can be realized by
taking $\mu=0$.

For the optimization \eqref{eq:JQoptMu},
the ISTA algorithm produces a sequences of estimates $\Qbf_k$, $k=0,1,2,\ldots$
 with updates that can be described as follows:
At each iteration $k$, we find an $\alpha_k > 0$ such that
\beqa
    \lefteqn{ J_{\mu}(\Qbf) \leq \Jbar_\mu(\Qbf,\Qbf_k) := J_\mu(\Qbf_k) } \nonumber \\
    && +
    \frac{\partial J_\mu(\Qbf_k)}{\partial \Qbf} \cdot
    (\Qbf - \Qbf_k) + \frac{1}{2 \alpha_k}\|\Qbf - \Qbf_k\|^2_F, \label{eq:Jbardef}
\eeqa
for all possible $\Qbf \geq 0$.  We will discuss how to select such a value $\alpha_k$
momentarily.  Thus, $\Jbar_\mu(\Qbf,\Qbf_k)$ represents a quadratic
upper bound on the true objective $J_\mu(\Qbf)$ that matches the true objective
at $\Qbf=\Qbf_k$.  In the case of the objective function \eqref{eq:JQ}, it is easy to
check that the derivative in any direction $\Delta$ is given by
\beq \label{eq:derivSk}
    \frac{\partial J_\mu(\Qbf_k)}{\partial \Qbf} \cdot \Delta = \Tr(\Sbf_k^* \Delta),
\eeq
where
\beq \label{eq:Skdef}
    \Sbf_k = \sum_{\ell=1}^L \left[ \frac{1}{\lambda_\ell(\Qbf_k)} -
    \frac{y_\ell}{\lambda_\ell^2(\Qbf_k)} \right] \ubf_\ell\ubf_\ell^* + \mu\Ibf.
\eeq
The concept in the ISTA algorithm is, at each iteration $k$, to minimize the
upper bound $\Jbar_\mu(\Qbf,\Qbf_k)$ instead of the true objective $J_\mu(\Qbf)$:
\beqa
    \lefteqn{ \Qbf_{\kp1} = \argmin_{\Qbf \geq 0} \Jbar_\mu(\Qbf,\Qbf_k) } \nonumber \\
    &\stackrel{(a)}{=}&
    \argmin_{\Qbf > 0} \Tr(\Sbf_k^*\Qbf) + \frac{1}{2\alpha_k}\|\Qbf-\Qbf_k\|^2_F \nonumber \\
    &\stackrel{(b)}{=}& \argmin_{\Qbf \geq 0} \|\Qbf - \Qbf_k + \alpha_k \Sbf_k\|^2_F \nonumber \\
    &\stackrel{(c)}{=}& T_+\left( \Qbf_k - \alpha_k \Sbf_k \right),
\eeqa
where in step (a) we substituted the definition of $\Jbar_\mu(\cdot)$ in \eqref{eq:Jbardef}
and derivative \eqref{eq:derivSk} and removed terms that do not depend on $\Qbf$;
step (b) follows from rearranging the quadratic and in step (c) the operator
$T_+(\Pbf)$ is called the \emph{proximal operator} and is given by
\beq \label{eq:Tplus}
    T_+(\Pbf) := \argmin_{\Qbf > 0} \|\Qbf-\Pbf\|^2_F.
\eeq
It is shown in \cite{cai2010singular}
that this minimization can be computed easily via an eigenvalue
decomposition.  Specifically, when $\Pbf=\Pbf^*$, we know that
$\Pbf = \Ubf\Dbf\Ubf^*$ for some unitary $\Ubf$ and real diagonal
$\Dbf = \diag(\dbf)$.  In this case, the proximal operator is
\[
    T_+(\Pbf) = \Ubf \diag\left[  \max(\dbf,0) \right] \Ubf^*,
\]
which simply thresholds the eigenvalues of the matrix.
The resulting algorithm is shown in Algorithm~\ref{algo:ISTA}.

\begin{algorithm}
\caption{ML Estimation via ISTA}
\begin{algorithmic}[1]  \label{algo:ISTA}
\REQUIRE{
Matrix search directions $\ubf_\ell$, power values $y_\ell$, $\ell=1,\ldots,L$,
and SNR $\gamma$.
}

\STATE{ $t \gets 0$  }
\STATE{ Initialize $\Qbf_0 \geq 0$  }
\REPEAT
    \STATE{ $\lambda_\ell \gets \ubf^*_\ell(\Qbf_k + \gamma^{-1}\Ibf)\ubf_\ell$, $\forall \ell$}
    \STATE{ Compute the gradient $\Sbf_k$ from \eqref{eq:Skdef} }
    \STATE{ Select step size $\alpha_k > 0$ }
    \STATE{ $\Qbf_{\kp1} \gets T_+(\Qbf_k - \alpha_k\Sbf_k)$ }
\UNTIL{Terminated}

\end{algorithmic}
\end{algorithm}

A key property of the ISTA algorithm is that the objective function monotonically decreases
for a sufficiently small step-size.

\begin{proposition} \label{prop:ISTAConv}
Consider the output of the ISTA algorithm, Algorithm \eqref{algo:ISTA}, generated for
a set of measurement vectors $\ubf_\ell$, power measurements $y_\ell$ and SNR value $\gamma > 0$.
Then, there exists a minimum step size $\alphabar > 0$ such that
if $\alpha_k < \alphabar$ for all $k$, the objective $J_\mu(\Qbf)$ monotonically
decreases.
\end{proposition}
\begin{IEEEproof}
From Taylor's Theorem, we know that the bound \eqref{eq:Jbardef} will be satisfied
if
\beq \label{eq:JHessBnd}
    \frac{\partial^2}{\partial \Qbf^2} J_{\mu}(\Qbf) \leq \frac{1}{2\alpha_k} \Ibf.
\eeq
Now, since $\Qbf \geq \zerobf$, the power levels $\lambda_\ell(\Qbf)$
in \eqref{eq:lamell} will be bounded
below by
\[
    \lambda_\ell(\Qbf) \geq \frac{\|\ubf_\ell\|^2}{\gamma}.
\]
Using this bound, one can verify that there is a global upper bound on the Hessian in
the left-hand side of \eqref{eq:JHessBnd}.  This implies that there exists an $\alphabar > 0$
such that if $\alpha_k < \alphabar$ then \eqref{eq:JHessBnd} will be satisfied and therefore
so will the bound \eqref{eq:Jbardef}.
We therefore have that at each iteration  $k$,
\beq
    J_\mu(\Qbf_{\kp1}) \stackrel{(a)}{\leq} \Jbar_\mu(\Qbf_{\kp1},\Qbf_k)
       \stackrel{(b)}{\leq} \Jbar_\mu(\Qbf_k,\Qbf_k)
       \stackrel{(c)}{=} J_\mu(\Qbf_k),
\eeq
where step (a) follows from quadratic upper bound approximation in \eqref{eq:Jbardef},
step (b) is based on monotonically decreasing behavior of the cost function when we are applying ISTA algorithm,
and in step (c) we substituted the $\Qbf_k$ in \eqref{eq:Jbardef}.
This shows that for sufficiently small step sizes, the objective function decreases monotonically.
\end{IEEEproof}

A more refined analysis along the lines of \cite{BeckTeb:09} will show additionally
that the $J_\mu(\Qbf_k)$ converges to a local minima, which will also be a global minima
since the function is convex.

\subsection{Adaptive Step-size Selection} \label{sec:adaptStep}
Proposition~\ref{prop:ISTAConv} guarantees that for a sufficiently small step-size $\alphabar$,
the algorithm is guaranteed to converge.  However, selecting a single step-size that works for all $\Qbf_k$
may require a very small value, slowing the rate of the algorithm.
We thus use a simple, standard backtracking adaptive step-size method
\cite{nocedal2006numerical} as follows.
In each iteration of our algorithm, we attempt
a candidate step size $\alpha_k > 0$ and compute a test estimate
$\tilde{\Qbf}_{\kp1}$.  We know that, from a first-order approximation
of the objective,
\[
    J_\mu(\tilde{\Qbf}_{\kp1}) \approx J_\mu(\Qbf_k) +
        \Tr(\Sbf_k^*(\tilde{\Qbf}_{\kp1}-\Qbf_{k})).
\]
We thus test the condition
\beq \label{eq:JdecCond}
J_\mu(\tilde{\Qbf}_{\kp1}) < J_\mu(\Qbf_k) + \rho
    \Tr(\Sbf_k^*(\tilde{\Qbf}_{\kp1}-\Qbf_{k})),
\eeq
for some parameter $\rho \in (0,1)$.  Typically, we take $\rho = 1/2$.
If the condition \eqref{eq:JdecCond} is met, we accept the
candidate by setting $\Qbf_{\kp1} = \tilde{\Qbf}_{\kp1}$ and
increase the step-size $\alpha_{\kp1} = 2\alpha_k$.  On the
other hand, if the condition \eqref{eq:JdecCond} fails,
we discard the candidate by setting $\Qbf_{\kp1} = \Qbf_k$ and
decrease the step-size $\alpha_{\kp1} = \alpha_k/2$.

\subsection{Approximate ML Estimation via a GLM}\label{sec:optimGLM}

The ISTA algorithm, Algorithm~\ref{algo:ISTA},
is conceptually simple.  But, the optimization may not be feasible
for real-time implementations.  The main challenge is the thresholding step $T_+(\cdot)$.
Each thresholding step requires a eigenvalue decomposition.  As we will see in the simulations,
the algorithm often needs 100 iterations.
The key insight is given by the following lemma:

\begin{lemma} \label{lem:optEquiv}
Given a set of measurement vectors $\ubf_\ell$ define the set
\beq \label{eq:Gdef}
    \Gbf = \Big\{\qbf\Big|\Qbf = \sum_{\ell=1}^L q_\ell \ubf_\ell \ubf_\ell^*+q_0\Ibf\geq0\Big\}.
\eeq
Then, the optimization \eqref{eq:JQoptMu} can be rewritten as
an optimization over $\qbf \in \Gbf$
via the equivalence
\beq \label{eq:Proof}
    \min_{\Qbf \geq \zerobf} \, J_\mu(\Qbf) = \min_{\qbf \in \Gbf} \,
        f(\Abf\qbf),
\eeq
where $\Abf$ is the matrix with components
\beq \label{eq:Adefine}
    A_{\ell j} = \begin{cases}
        N & \mbox{ if } j = 0, \ell = 0  \\
        \|\ubf_j\|^2 & \mbox{ if } j = 0, \ell > 0 \\
        \|\ubf_\ell\|^2 & \mbox{ if } \ell = 0, j > 0 \\
        |\ubf_\ell^*\ubf_j|^2 & \mbox{ if } j,\ell=1,\ldots,L,
        \end{cases}
\eeq
and $f(\zbf) = \sum_{\ell=0}^L f_\ell(z_\ell)$, and
\begin{subequations} \label{eq:fdef}
\beqa
    f_0(z_0) &=& \mu z_0 \\
    f_\ell(z_\ell) &=& \log\left(z_\ell+\frac{1}{\gamma}\|\ubf_\ell\|^2\right)+
        \frac{y_\ell}{z_\ell+\frac{1}{\gamma}\|\ubf_\ell\|^2}, \nonumber \\
        & & ~ \ell=1,\ldots,L.
\eeqa
\end{subequations}
\end{lemma}
\begin{IEEEproof}  See Appendix \ref{sec:appendix}. \end{IEEEproof}

To understand the purpose of this lemma, recall that the chief computational difficulty
in the ISTA algorithm arises from the thresholding step to impose the positivity constraint
on $\Qbf \geq \zerobf$.
The above lemma shows that the optimization \eqref{eq:JQoptMu}
over $N \x N$ matrices $\Qbf \geq \zerobf$ can be replaced
by an optimization over an $(L+1)$-dimensional vector $\qbf \in \Gbf$.

Unfortunately, the constraint in $\Gbf$ in \eqref{eq:Gdef} still requires
a positivity constraint on the resulting matrix $\Qbf$.
However, this problem
can be approximately circumvented as follows:
We know that if $\qbf > 0$ then
\[
    \Qbf = \sum_{\ell=1}^L q_\ell \ubf_\ell \ubf_\ell^*+q_0\Ibf\geq \zerobf
    \Rightarrow \qbf \in \Gbf.
\]
The converse is not true.  That is, it is not the case that $\qbf \in \Gbf$ implies
that $\qbf \geq 0$.  However, instead of searching over all $\qbf \in \Gbf$, we can
replace the optimization in \eqref{eq:Proof} with the approximate ML optimization
\beq \label{eq:JQopt1}
    \qbfhat = \argmin_{\qbf \geq 0} f(\Abf\qbf).
\eeq

The resulting optimization \eqref{eq:JQopt1}
has a particularly simple structure.
First, the vector $\qbf$ has only componentwise constraints:
$\qbf \geq 0$.
Second, for any $\qbf$,
the objective function in \eqref{eq:JQopt1} can be evaluated via
a linear transform $\zbf = \Abf\qbf$ and followed by a sum
of convex functions \eqref{eq:fdef} on the components $z_\ell$.

This type of optimization described by a separable function of a linear
transform of the vector arises in a wide range of applications.
The most common application is in inference problems for
so-called generalized linear models (GLMs)~\cite{NelWed:72}.

\subsection{An ISTA Algorithm for the Approximate ML}
As before, we can apply an ISTA-type approach to the optimization
\eqref{eq:JQopt1} as follows.
Let $\qbf_k$ be the estimate at the $k$-th iteration and
be its $\zbf_k = \Abf\qbf_k$ be its transform.
We then find a $\alpha_k > 0$ such that,
\beqa
    \lefteqn{ f(\Abf\qbf) \leq \fbar(\qbf,\qbf_k) := f(\Abf\qbf_k) } \nonumber \\
    && +
    \frac{\partial f(\Abf\qbf_k)}{\partial \qbf} \cdot
    (\qbf - \qbf_k) + \frac{1}{2 \alpha_k}\|\qbf - \qbf_k\|^2_2, \label{eq:fbardef}
\eeqa
for all possible $\qbf\geq 0$.
So, $\fbar(\qbf,\qbf_k)$ is a
quadratic upper bound on the true objective $f(\Abf\qbf)$ that
matches the true objective at the current estimate $\qbf=\qbf_k$.
Also, the derivative in \eqref{eq:fbardef} is given by
\beq \label{eq:derivq}
    \frac{\partial f(\Abf\qbf)}{\partial \qbf}\Bigr|_{\qbf=\qbf_k} = \sbf_k^*,
        \quad
    \sbf_k := \Abf^*\frac{\partial f(\zbf_k)}{\partial \zbf}.
\eeq
Then, as before, at each iteration $k$, the ISTA algorithm minimizes the
upper bound $\fbar(\qbf,\qbf_k)$ instead of the true objective
$f(\Abf\qbf)$:
\beqa
    \lefteqn{\qbf_{\kp1} = \argmin_{\qbf \geq 0} \fbar(\qbf,\qbf_k) }
        \nonumber \\
    &\stackrel{(a)}{=}&
    \argmin_{\qbf > 0} \sbf_k^*\qbf + \frac{1}{2\alpha_k}\|\qbf-\qbf_k\|^2_2+\mu\vbf^T\qbf \nonumber \\
    &\stackrel{(b)}{=}& \argmin_{\qbf \geq 0} \|\qbf- \qbf_k + \alpha_k \sbf_k\|^2_2+\mu\vbf^T\qbf \nonumber \\
    &\stackrel{(c)}{=}& T_+\left( \qbf_k - \alpha_k \sbf_k  \right),
\eeqa
where in step (a) we substituted the definition of
$\fbar(\cdot)$ in \eqref{eq:fbardef} and derivative \eqref{eq:derivq} and
removed terms that do not depend on $\qbf$;
step (b) follows from rearranging the quadratic and
in step (c) the operator $T_+(\pbf)$ is the proximal operator given by
\beq \label{eq:Tplus1}
    T_+(\pbf) := \argmin_{\qbf > 0} \|\qbf-\pbf\|^2_2.
\eeq
which is simply the quadratic approximation and
removing the negative components.
It is easily checked that the proximal operator \eqref{eq:Tplus1}
is given by a simple thresholding operation
\beq \label{eq:Tplus2}
    T_+(\pbf) := \max\{\pbf, 0\},
\eeq
which simply sets all negative components of $\pbf$ to zero.
The resulting algorithm is shown in Algorithm~\ref{algo:ISTAGLM}.

\begin{algorithm}
\caption{Approximate ML Estimation via ISTA}
\begin{algorithmic}[1]  \label{algo:ISTAGLM}
\REQUIRE{
Matrix search directions $\ubf_\ell$ and power values
$y_\ell$, $\ell=1,\ldots,L$, and SNR $\gamma$.
}

\STATE{ Construct $\Abf$ in \eqref{eq:Adefine}}
\STATE{ $k \gets 0$  }
\STATE{ Initialize $\qbf_0 \geq 0$  }
\REPEAT
    \STATE{ $\zbf_k \gets \Abf\qbf_k$}
    \STATE{ $\sbf_k \gets \Abf^*\partial f(\zbf_k)/\partial \zbf$}
    \STATE{ Select step size $\alpha_k > 0$ }
    \STATE{ $\qbf_{\kp1} \gets \max\{0, \qbf_k-\alpha_k \sbf_k \}$ }
\UNTIL{Terminated}

\end{algorithmic}
\end{algorithm}

The main advantage of the Approximate ML algorithm,
Algorithm~\ref{algo:ISTAGLM} in comparison to Algorithm~\ref{algo:ISTA}
is its complexity.  Each iteration involves only multiplications by $\Abf$
and $\Abf^*$ as well as simple scalar derivatives.
In particular, unlike Algorithm~\ref{algo:ISTA},
there are no eigenvalue value decompositions needed for the thresholding
step.

Also, as with Algorithm~\ref{algo:ISTA},
the objective function  monotonically decreases
with sufficiently small step-sizes $\alpha_k$.
Specifically, suppose that at some time $k$,
$\alpha_k$ is sufficiently small that
the bound \eqref{eq:fbardef} is satisfied for all $\qbf$.
Then, we have
\beq
    \fbar(\qbf_{\kp1}) \stackrel{(a)}{\leq} \fbar(\qbf_{\kp1},\qbf_k)
       \stackrel{(b)}{\leq} \fbar(\qbf_k,\qbf_k)
       \stackrel{(c)}{=} f(\qbf_k),
\eeq
where step (a) follows from quadratic upper bound approximation
in \eqref{eq:fbardef}, step (b) is based on monotonically
decreasing behavior of the cost function when we are applying
ISTA algorithm, and in step (c)
we substituted the $\qbf_k$ in \eqref{eq:fbardef}.
As in Section~\ref{sec:adaptStep}, we can adapt the step-size
with a backtracking type rule.

\section{Numerical Simulation}

\subsection{Single-path Channel}

To assess the performance of the proposed estimators,
we first consider a theoretical single path channel.
Specifically, we assume that, in each measurement $\ell$
and transmission $d$, the
single-input multi-output (SIMO) channel is given by
\beq \label{eq:hsing}
    \hbf_{\ell d} = \alpha_{\ell d} \vbf(\theta, \phi),
\eeq
where $\theta$ and $\phi$ are the horizontal and vertical
angles of arrival (AoAs) of the path,
$\vbf(\theta,\phi)$ is the vector antenna response
to the path, and $\alpha_{\ell d}$ is a complex scalar representing
the small scale fading -- see \cite{TseV:07} for details.
The parameters $\theta$ and $\phi$ are determined by
the large-scale path directions and are thus assumed to be constant.
However, we assume that the small-scale parameter $\alpha_{\ell d}$
is independently Rayleigh faded across different measurements
$\ell$ and $d$.  Under this single-path model,
the average spatial covariance is then given by the rank one matrix
\beq \label{eq:Qsing}
    \Qbf = \Exp\left[ \hbf_{\ell d}\hbf_{\ell d}^* \right] =
        \Exp|\alpha|^2 \vbf(\theta, \phi)\vbf^*(\theta, \phi).
\eeq
We assume the power is normalized so that $\Exp|\alpha|^2=1$.
Following \cite{AkdenizCapacity:14}, we assume
a two-dimensional $4 \x 4$ $\lambda/2$ uniform linear array.
This array size can be easily accommodated in a mobile in the mmW range.
For example, at 28~GHz, the array would be only approximately 1.5 cm$^2$.
We set the SNR to 10 dB per antenna.

We then simulate the algorithm through 1000 Monte Carlo trials.
In each trial, we generate random AoAs $(\theta,\phi)$
and random search directions $\ubf_\ell$ for the $L$ measurements.
The number of measurements $L$ is varied.
The random search directions $\ubf_\ell$ are taken as the
antenna response along random angles that are generated in an i.i.d.\
manner.
Following \cite{barati2014directional},
we also take the diversity order $D=4$.
We then run the ML and approximate ML algorithm to compute
estimates $\Qbfhat$ of the true spatial covariance matrix $\Qbf$.

To evaluate the accuracy of the estimate $\Qbfhat$, we measure the loss
in beamforming resulting from the estimation errors.
In general, given the true spatial covariance matrix $\Qbf$,
the optimal long-term beamforming vector $\wbf_{opt}$ is the unit
vector directed along the maximal eigenvector of $\Qbf$.
The optimal long-term beamforming gain is then
\[
    G_{opt} := \wbf_{opt}^*\Qbf\wbf_{opt} = \lambda_{max}(\Qbf),
\]
where $\lambda_{max}(\Qbf)$ is the maximal eigenvalue of $\Qbf$.
For a rank-one single-path channel, the optimal beamforming
vector is simply the vector aligned to the receive spatial signature,
$\wbf_{opt} \propto \vbf(\theta,\phi)$.  Also, assuming the
spatial covariance matrix is normalized to unity $\Tr(\Qbf)=1$,
the optimal beamforming gain
is simply $N$, the dimension of the antenna array.
See \cite{Lozano:07} for details.

To evaluate the loss from channel estimation errors, we suppose that
the receiver applies a beamforming gain from the estimated
covariance matrix $\Qbfhat$.  That is, we compute $\wbfhat$ from the
maximal eigenvector of $\Qbfhat$ and then compute the actual gain,
\[
    G := \wbfhat^*\Qbf\wbfhat.
\]
The loss is then given by
\[
    \mbox{loss} = 10\log_{10}(G_{opt}/G),
\]
which is the loss (in dB) due to the channel estimation error.

\begin{figure}
\centering {\includegraphics[trim=0.1cm 6cm 0.1cm 6cm ,clip=true, width=1\linewidth]{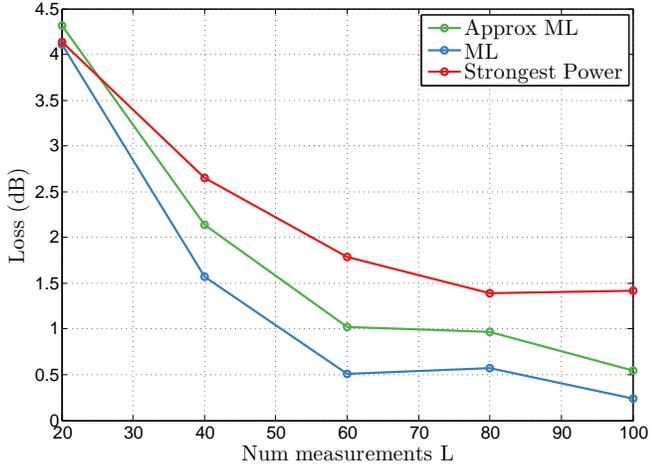}}
\caption{\textbf{Algorithm performance on a single-path channel.}
To estimate the performance of the algorithms,
we computed the optimal beamforming vector for the estimated
channel covariance matrix and then measured the loss in beamforming
gain from applying the estimated vector relative to the optimal vector.
The loss is plotted as a function of the number of power measurements $L$
for different estimation algorithms assuming an ideal single
path channel and a per antenna SNR of 10~dB. }
\label{fig:LossvsNwSingle}
\end{figure}

Fig.~\ref{fig:LossvsNwSingle} plots the mean value of the loss
as a function of the number of measurements $L$.
There are several points to observe.
First, we observe that with $L$ around 60 measurements,
the exact ML algorithm
obtains a loss of less than 0.5 dB.  Second, tt should be pointed out that
since the antenna array has dimension $N=16$, the Hermetian matrix $\Qbf$
has $N(N+1)/2 = 136$ unknowns.  Hence, the low-rank method is successful
in estimating the matrix well even though the number of measurements is below
the number of free parameters.  This property is precisely the value
of the non-negative constraints.
Third, the approximate ML method is only slightly inferior to the exact
method.  As mentioned above, the approximate ML method is
significantly simpler to implement and thus the small additional loss
may justify its use.

Finally, as a point of comparison, Fig.~\ref{fig:LossvsNwSingle},
plots the beamforming gain from a simple algorithm based on
selecting the beamforming direction that resulted in the maximum power.
Interestingly, this simple and intuitive algorithm performs considerably
worse than the proposed method.
For example, at $L=60$ measurements, the loss is 1.5 dB, about 1 dB worse
than the proposed ML estimation method.

\subsection{Multi-path Channel using NYC Measurements}
An important and surprising finding of the mmW measurements in
New York City reported in
\cite{Rappaport:12-28G,Samimi:AoAD,rappaportmillimeter} is
that in urban micro-cellular type deployments,
mmW signals are likely to propagate via multiple paths to
the receiver.  Although mmW signals are blocked by many
materials, many street-level locations were able to see base stations
at 100 to 200m via diffuse scattering and reflections, even
when situated in non-line-of-sight (NLOS) locations.
It is precisely this phenomena that enables mmW pico and micro-cellular
type deployments.

To validate the channel estimation
algorithms in these scenarios, we
next simulated the algorithms with the spatial covariance
matrix $\Qbf$ generated from the model \cite{AkdenizCapacity:14}
derived from the New York City measurements
\cite{Rappaport:12-28G,Samimi:AoAD,rappaportmillimeter}
made at 28~GHz.   The model in \cite{AkdenizCapacity:14}
follows a similar form to the standard 3GPP / ITU
model~\cite{3GPP36.814,ITU-M.2134} with parameters fit to the
mmW measurements.  Specifically, the channel is composed of a random
number of clusters, each cluster having some random angular spread and power.
Based on data analysis in \cite{AkdenizCapacity:14},
the mmW channel typically has one to three clusters
with a small angular spread in each clusters.  Details can be found in
\cite{AkdenizCapacity:14}.

\begin{figure}
\centering {\includegraphics[trim=0.1cm 6cm 0.1cm 6cm ,clip=true, width=1\linewidth]{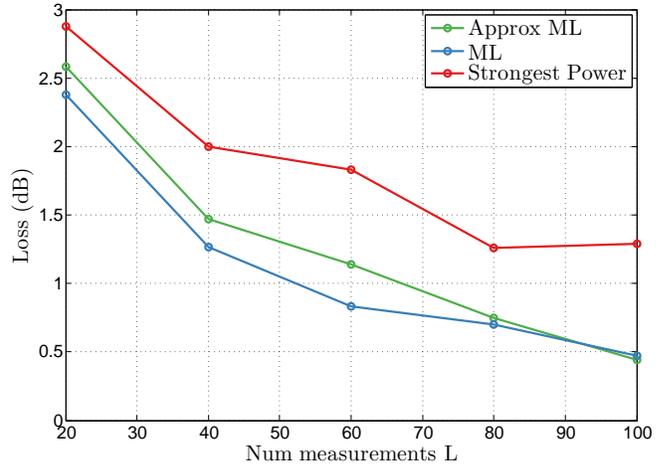}}
\caption{\textbf{Algorithm performance on a realistic multi-path channel.}
Details are identical to Fig.~\ref{fig:LossvsNwSingle}
except we use a multi-path channel model from \cite{AkdenizCapacity:14}
based on the real NYC measurements at 28~GHz~\cite{rappaportmillimeter}.} 
\label{fig:LossvsNwMulti}
\end{figure}

Fig. ~\ref{fig:LossvsNwMulti} plots the loss for different
number of directions, $L \in \{20,60,80,100\}$.  In comparison
to the single path case, we see we need slightly more measurements.
For example, for a 0.5 dB loss, we need $L=100$ measurements.
This number is still less than the number of free parameters.
However, the other features remain the same.
Specifically, the approximate ML results in only a small loss
relative to the exact ML and both methods are considerably better than
the simple strongest power method.

\subsection{Tuning the sparsity factor $\mu$}
\begin{figure}
\centering {\includegraphics[trim=0.1cm 6cm 0.1cm 6cm ,clip=true, width=1\linewidth]{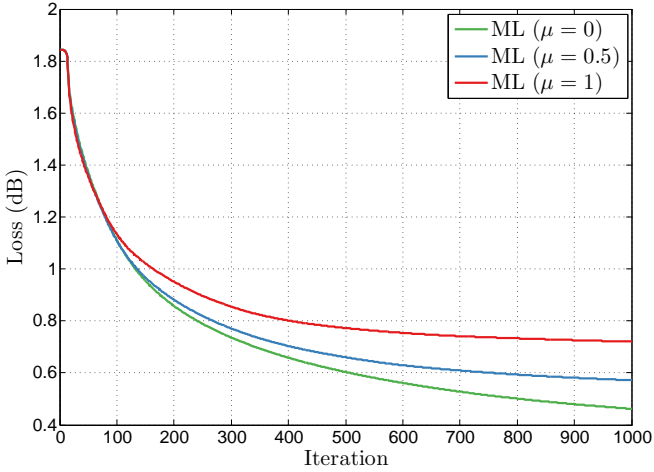}}
\caption{Calculated loss for single path channel versus iterations that ISTA has been run for 50 different beamforming directions and different sparsity factors, $\mu = \{0,\,0.5,\,1\}$}
\label{fig:SparsityFactor}
\end{figure}

In the simulations up to now, we have set the sparsity regularization
parameter $\mu=0$.  That is, we have used the unregularized
ML objective \eqref{eq:JQ} instead of the regularized
objective \eqref{eq:JQmu}.
However, given the low-rank nature of the channel, one may expect that
adding a regularization term to force sparsity in the singular values
of $\Qbf$ would improve the estimation.
To test this hypothesis, we evaluated the beamforming loss
as a function of $\mu$.  Fig.~\ref{fig:SparsityFactor} shows
the average beamforming loss at $L=50$ measurements as a function of the
iteration number for three different values of $\mu$.
It can be seen that using a non-zero value of $\mu$ only makes the
performance worse.
Similar results were found at different values of $L$ as well.
Also, for the multipath channel, using $\mu > 0$ was even worse
since the channel is inherently higher rank.

As described in Section~\ref{sec:sparseReg}, the fact that $\mu=0$
is optimal is not entirely surprising.  The optimization \eqref{eq:JQopt}
already imposes the positivity constraint $\Qbf \geq \zerobf$.
Similar to \cite{lee2001algorithms},
non-negative constraints tend to naturally impose sparsity,
so it is possible that additional sparsity regularization is not necessary.

\section*{Conclusions}

Millimeter wave systems rely centrally on directional transmissions.
Due to the rapid variations in the channel and need for low-latency
communication, algorithms for fast spatial channels will thus be
key for the successful deployment of these technologies.
In this work, we have considered the estimation of the long-term
receiver-side spatial covariance of the channel from analog beamformed
power measurements.  ML estimation is shown to be equivalent to
an optimization that appears as a noisy, non-negative matrix
completion problems.  Fast algorithms were developed to
solve this optimization and were demonstrated on both ideal
single path channels as well as channel models derived from
real measurements in urban deployments.
The algorithms show relatively fast convergence (~100 iterations)
and can provide good tracking with significantly less number of
measurements than unknowns.

Several future avenues of work are possible.  First,
we have considered only analog beamforming.
Low-bit, fully digital, as proposed in
\cite{Madhow:ADC,Madhow:largeArray}, may offer significantly
improved performance and should be investigated.  Also,
the current algorithms assumes the long-term statistics
are constant.  Future work may also consider tracking of these
parameters.  Finally, the number of iterations for convergence
is still somewhat long.  Other approaches including Fast ISTA
\cite{Nesterov:07} and approximate message passing methods
\cite{Rangan:11-ISIT} may also be considered.

\appendix

\section*{Proof of Lemma~\ref{lem:optEquiv}} \label{sec:appendix}
First, we show that
\beq \label{eq:Jequiv}
    \min_{\Qbf \geq \zerobf} J_\mu(\Qbf) =
    \min_{\Qbf \in \Gbf} J_\mu(\Qbf).
\eeq
Since $\Gbf$ in \eqref{eq:Gdef} consists of positive matrices,
we know that the left-hand side of \eqref{eq:Jequiv} is less than
or equal to the right-hand side.  Hence, we need to show that
the right-hand side is less than or equal to the left-hand side.
To prove this, we need to show the following:
For any $\Qbf \geq \zerobf$, there exists a $\Qbf_0 \in \Gbf$
with $J_\mu(\Qbf_0) = J_\mu(\Qbf)$.   Thus, let $\Qbf \geq \zerobf$.
Decompose $\Qbf$ as
\[
    \Qbf = \Qbf_0 + \Rbf,
\]
where $\Qbf_0 \in \Gbf$ and $\Rbf \bot \Gbf$, where the orthogonality
is with respect to the standard inner product between matrices,
$\Abf \cdot \Bbf = \Tr(\Abf^*\Bbf)$.
Now, since $\Rbf \bot \Gbf$ we have
\beqan
    && \Rbf ~\bot~ \Ibf \Rightarrow \Tr(\Rbf) = 0, \\
    && \Rbf ~\bot ~\ubf_\ell \ubf_\ell^* \Rightarrow \ubf_\ell^*\Rbf\ubf_\ell = 0.
\eeqan
Also, since $\Qbf_0 \in \Gbf$, we have that
\[
    \Qbf_0 =  \sum_{\ell=1}^L q_\ell \ubf_\ell \ubf_\ell^*+q_0\Ibf,
\]
for some coefficients $\qbf$.
So we have,
\beqa
 \lambda_\ell(\Qbf) &=& \ubf_\ell^*
    \left[\Qbf_0 + \Rbf + \gamma^{-1} \Ibf\right]\ubf_\ell \nonumber\\
 &=& \ubf_\ell^*\left[\Qbf_0 + \gamma^{-1} \Ibf\right]\ubf_\ell \quad \Big(\Rbf \bot \ubf_\ell \ubf_\ell^*\Big) \nonumber\\
 &=& \lambda_\ell(\Qbf_0) \label{eq:lamQeq}
\eeqa
Hence, from \eqref{eq:JQ}, $J(\Qbf)=J(\Qbf_0)$.
Also, since $\Tr(\Rbf) = 0$,
\beq \label{eq:TrQeq}
    \Tr(\Qbf) = \Tr(\Qbf_0+\Rbf) = \Tr(\Qbf_0).
\eeq
Therefore, from \eqref{eq:JQmu}, $J_\mu(\Qbf) = J_\mu(\Qbf_0)$.
Hence, for any $\Qbf \geq \zerobf$, we can find a $\Qbf_0 \in \Gbf$
with the same objective value $J_\mu(\Qbf) = J_\mu(\Qbf_0)$.
This proves \eqref{eq:Jequiv}.

To prove \eqref{eq:Proof}, we need to show
\beq \label{eq:Jmuf}
    J_\mu(\Qbf) = f(\Abf\qbf),
\eeq
whenever $\Qbf$ is of the form
\beq \label{eq:Qsum}
    \Qbf = \sum_{\ell=1}^L q_\ell \ubf_\ell \ubf_\ell^*+q_0\Ibf.
\eeq
Let $\zbf=\Abf\qbf$.  First observe that
\beqa
    \lefteqn{ \lambda_\ell(\Qbf) \stackrel{(a)}{=}
       \ubf_\ell^*(\Qbf+ \gamma^{-1}\Ibf)\ubf_\ell } \nonumber \\
     &\stackrel{(b)}{=}& \sum_{j=1}^L q_j |\ubf_j^*\ubf_\ell|^2 + (q_0+ \gamma^{-1})|\ubf_\ell|^2 \nonumber\\
     &\stackrel{(c)}{=}& \sum_{j=0}^L A_{\ell j}q_j + \gamma^{-1}\|\ubf_\ell\|^2
     \stackrel{(d)}{=} z_\ell + \gamma^{-1}\|\ubf_\ell\|^2,
\eeqa
where (a) follows from \eqref{eq:lamell},
(b) follows from \eqref{eq:Qsum},
(c) follows from the definition of the matrix components in
\eqref{eq:Adefine} and (d) follows from the fact that
$z_\ell = (\Abf\qbf)_\ell$.
Hence, the objective function $J(\Qbf)$ in \eqref{eq:JQ}
is given by
\beqa
    J(\Qbf) &=& \sum_{\ell=1}^L \log\left[
    z_\ell + \frac{1}{\gamma}\|\ubf_\ell\|^2
        \right] + \frac{y_\ell}{z_\ell + \frac{1}{\gamma}\|\ubf_\ell\|^2}
        \nonumber \\
        &=& \sum_{\ell=1}^L f_\ell(z_\ell), \hspace{2cm} \label{eq:Jfell}
\eeqa
where the last step follows from \eqref{eq:fdef}.
Also using \eqref{eq:Qsum},
\beqa
    \lefteqn{ \Tr(\Qbf) = Nq_0 + \sum_{j=1}^L q_\ell \|\ubf_\ell\|^2 }
    \nonumber \\
    &\stackrel{(a)}{=} \sum_{j=0}^L A_{0j}q_j = z_0,
\eeqa
where in (a) we again used \eqref{eq:Adefine}.
Hence, from \eqref{eq:fdef},
\beq \label{eq:Jf0}
    \mu\Tr(\Qbf) = f_0(z_0).
\eeq
Combining \eqref{eq:JQmu}, \eqref{eq:Jfell} and \eqref{eq:Jf0},
we see that
\beq
    J_\mu(\Qbf) = \sum_{\ell=0}^L f_\ell(z_\ell)
    = f(\Abf\qbf).
\eeq
This proves \eqref{eq:Jmuf} and the proof is complete.

\bibliographystyle{IEEEtran}
\bibliography{bibl}

\newcommand{\SortNoop}[1]{}
\begin{thebibliography}{10}
\providecommand{\url}[1]{#1}
\csname url@samestyle\endcsname
\providecommand{\newblock}{\relax}
\providecommand{\bibinfo}[2]{#2}
\providecommand{\BIBentrySTDinterwordspacing}{\spaceskip=0pt\relax}
\providecommand{\BIBentryALTinterwordstretchfactor}{4}
\providecommand{\BIBentryALTinterwordspacing}{\spaceskip=\fontdimen2\font plus
\BIBentryALTinterwordstretchfactor\fontdimen3\font minus
  \fontdimen4\font\relax}
\providecommand{\BIBforeignlanguage}[2]{{%
\expandafter\ifx\csname l@#1\endcsname\relax
\typeout{** WARNING: IEEEtran.bst: No hyphenation pattern has been}%
\typeout{** loaded for the language `#1'. Using the pattern for}%
\typeout{** the default language instead.}%
\else
\language=\csname l@#1\endcsname
\fi
#2}}
\providecommand{\BIBdecl}{\relax}
\BIBdecl

\bibitem{CiscoVNI:latest}
Cisco, ``{Cisco Visual Network Index}: Global mobile traffic forecast update,''
  2013.

\bibitem{KhanPi:11-CommMag}
F.~Khan and Z.~Pi, ``{An introduction to millimeter-wave mobile broadband
  systems},'' \emph{IEEE Comm. Mag.}, vol.~49, no.~6, pp. 101 -- 107, Jun.
  2011.

\bibitem{PietBRPC:12}
P.~Pietraski, D.~Britz, A.~Roy, R.~Pragada, and G.~Charlton, ``Millimeter wave
  and terahertz communications: Feasibility and challenges,'' \emph{ZTE
  Communications}, vol.~10, no.~4, pp. 3--12, Dec. 2012.

\bibitem{rappaportmillimeter}
T.~S. Rappaport, S.~Sun, R.~Mayzus, H.~Zhao, Y.~Azar, K.~Wang, G.~N. Wong,
  J.~K. Schulz, M.~Samimi, and F.~Gutierrez, ``{Millimeter Wave Mobile
  Communications for 5G Cellular: It Will Work!}'' \emph{IEEE Access}, vol.~1,
  pp. 335--349, May 2013.

\bibitem{RanRapE:14}
S.~Rangan, T.~S. Rappaport, and E.~Erkip, ``Millimeter-wave cellular wireless
  networks: Potentials and challenges,'' \emph{Proceedings of the IEEE}, vol.
  102, no.~3, pp. 366--385, March 2014.

\bibitem{BocHLMP:14}
F.~Boccardi, R.~W. Heath, A.~Lozano, T.~L. Marzetta, and P.~Popovski, ``Five
  disruptive technology directions for {5G},'' to appear in \emph{IEEE Comm.
  Magazine}, 2014.

\bibitem{Rappaport2014-mmwbook}
T.~S. Rappaport, R.~W. {Heath Jr.}, R.~C. Daniels, and J.~N. Murdock,
  \emph{Millimeter Wave Wireless Communications}.\hskip 1em plus 0.5em minus
  0.4em\relax Pearson Education, 2014.

\bibitem{Rappaport:02}
T.~S. Rappaport, \emph{Wireless Communications: Principles and Practice},
  2nd~ed.\hskip 1em plus 0.5em minus 0.4em\relax Upper Saddle River, NJ:
  Prentice Hall, 2002.

\bibitem{AkdenizCapacity:14}
M.~Akdeniz, Y.~Liu, M.~Samimi, S.~Sun, S.~Rangan, T.~Rappaport, and E.~Erkip,
  ``Millimeter wave channel modeling and cellular capacity evaluation,''
  \emph{IEEE J. Sel. Areas Comm.}, vol.~32, no.~6, pp. 1164--1179, June 2014.

\bibitem{Lozano:07}
A.~Lozano, ``Long-term transmit beamforming for wireless multicasting,'' in
  \emph{Proc.\ ICASSP}, vol.~3, 2007, pp. III--417--III--420.

\bibitem{KhanPi:11}
F.~Khan and Z.~Pi, ``{Millimeter-wave {M}obile {B}roadband ({MMB}):
  {U}nleashing 3-300GHz Spectrum},'' in \emph{Proc.\ IEEE Sarnoff Symposium},
  Mar. 2011.

\bibitem{KohReb:07}
K.-J. Koh and G.~M. Rebeiz, ``{0.13- m CMOS phase shifters for X-, Ku- and
  K-band phased arrays},'' \emph{IEEE J. Solid-State Circuts}, vol.~42, no.~11,
  pp. 2535--2546, Nov. 2007.

\bibitem{KohReb:09}
------, ``{A Millimeter-Wave (40–45 GHz) 16-Element Phased-Array Transmitter
  in 0.18-m SiGe BiCMOS Technology},'' \emph{IEEE J. Solid-State Circuts},
  vol.~44, no.~5, pp. 1498--1509, May 2009.

\bibitem{GuanHaHa:04}
X.~Guan, H.~Hashemi, and A.~Hajimiri, ``{A fully integrated 24-GHz
  eight-element phased-array receiver in silicon},'' \emph{IEEE J. Solid-State
  Circuts}, vol.~39, no.~12, pp. 2311--2320, Dec. 2004.

\bibitem{Heath:partialBF}
A.~Alkhateeb, O.~E. Ayach, G.~Leus, and J.~Robert W.~Heath, ``Hybrid
  analog-digital beamforming design for millimeter wave cellular systems with
  partial channel knowledge,'' in \emph{Proc.\ Information Theory and
  Applications Workshop (ITA)}, Feb. 2013.

\bibitem{Rappaport:12-28G}
Y.~Azar, G.~N. Wong, K.~Wang, R.~Mayzus, J.~K. Schulz, H.~Zhao, F.~Gutierrez,
  D.~Hwang, and T.~S. Rappaport, ``28 {GHz} propagation measurements for
  outdoor cellular communications using steerable beam antennas in {N}ew {Y}ork
  {C}ity,'' in \emph{Proc.\ IEEE ICC}, 2013.

\bibitem{Samimi:AoAD}
M.~Samimi, K.~Wang, Y.~Azar, G.~N. Wong, R.~Mayzus, H.~Zhao, J.~K. Schulz,
  S.~Sun, F.~Gutierrez, and T.~S. Rappaport, ``28 {GHz} angle of arrival and
  angle of departure analysis for outdoor cellular communications using
  steerable beam antennas in {N}ew {Y}ork {C}ity,'' in \emph{Proc. IEEE VTC},
  2013.

\bibitem{molisch2014propagation}
A.~F. M{\"o}lisch and F.~Tufevsson, ``Propagation channel models for
  next-generation wireless communications systems,'' \emph{IEICE Trans.
  Communications}, vol.~97, no.~10, pp. 2022--2034, 2014.

\bibitem{martone1998adaptive}
M.~Martone, ``An adaptive algorithm for antenna array low-rank processing in
  cellular {TDMA} base stations,'' \emph{IEEE Trans. Communications}, vol.~46,
  no.~5, pp. 627--643, 1998.

\bibitem{ottersten1996array}
B.~Ottersten, ``Array processing for wireless communications,'' in \emph{Proc.\
  IEEE Signal Processing Workshop on Statistical Signal and Array
  Processing}.\hskip 1em plus 0.5em minus 0.4em\relax IEEE, 1996, pp. 466--473.

\bibitem{wang1998blind}
X.~Wang and H.~V. Poor, ``Blind multiuser detection: A subspace approach,''
  \emph{IEEE Trans. Information Theory}, vol.~44, no.~2, pp. 677--690, 1998.

\bibitem{wright2009robust}
J.~Wright, A.~Ganesh, S.~Rao, Y.~Peng, and Y.~Ma, ``Robust principal component
  analysis: Exact recovery of corrupted low-rank matrices via convex
  optimization,'' in \emph{Proc.\ NIPS}, 2009, pp. 2080--2088.

\bibitem{lin2010augmented}
Z.~Lin, M.~Chen, and Y.~Ma, ``The augmented lagrange multiplier method for
  exact recovery of corrupted low-rank matrices,'' \emph{arXiv preprint
  arXiv:1009.5055}, 2010.

\bibitem{koltchinskii2011nuclear}
V.~Koltchinskii, K.~Lounici, A.~B. Tsybakov \emph{et~al.}, ``Nuclear-norm
  penalization and optimal rates for noisy low-rank matrix completion,''
  \emph{Annals of Statistics}, vol.~39, no.~5, pp. 2302--2329, 2011.

\bibitem{keshavan2010matrix}
R.~H. Keshavan, A.~Montanari, and S.~Oh, ``Matrix completion from a few
  entries,'' \emph{Information Theory, IEEE Transactions on}, vol.~56, no.~6,
  pp. 2980--2998, 2010.

\bibitem{rangan2012iterative}
S.~Rangan and A.~K. Fletcher, ``Iterative estimation of constrained rank-one
  matrices in noise,'' in \emph{Proc.\ IEEE ISIT}, 2012, pp. 1246--1250.

\bibitem{chenrobust}
Y.~Chen, Y.~Chi, and A.~J. Goldsmith, ``Robust and universal covariance
  estimation from quadratic measurements via convex programming,'' in
  \emph{Proc.\ ISIT}, Honolulu,HI, July 2014.

\bibitem{BeckTeb:09}
A.~Beck and M.~Teboulle, ``A fast iterative shrinkage-thresholding algorithm
  for linear inverse problem,'' \emph{SIAM J.\ Imag.\ Sci.}, vol.~2, no.~1, pp.
  183–--202, 2009.

\bibitem{cai2010singular}
J.-F. Cai, E.~J. Cand{\`e}s, and Z.~Shen, ``A singular value thresholding
  algorithm for matrix completion,'' \emph{SIAM Journal on Optimization},
  vol.~20, no.~4, pp. 1956--1982, 2010.

\bibitem{NelWed:72}
J.~A. Nelder and R.~W.~M. Wedderburn, ``Generalized linear models,'' \emph{J.\
  Royal Stat.\ Soc. Series A}, vol. 135, pp. 370--385, 1972.

\bibitem{Rappaport:28NYCPenetrationLoss}
H.~Zhao, R.~Mayzus, S.~Sun, M.~Samimi, J.~K. Schulz, Y.~Azar, K.~Wang, G.~N.
  Wong, F.~Gutierrez, and T.~S. Rappaport, ``28 {GHz} millimeter wave cellular
  communication measurements for reflection and penetration loss in and around
  buildings in {N}ew {Y}ork {C}ity,'' in \emph{Proc. IEEE ICC}, 2013.

\bibitem{barati2014directional}
C.~N. Barati, S.~A. Hosseini, S.~Rangan, P.~Liu, T.~Korakis, and S.~S. Panwar,
  ``Directional cell search for millimeter wave cellular systems,'' \emph{arXiv
  preprint arXiv:1404.5068}, 2014.

\bibitem{TseV:07}
D.~Tse and P.~Viswanath, \emph{{Fundamentals of Wireless Communication}}.\hskip
  1em plus 0.5em minus 0.4em\relax Cambridge University Press, 2007.

\bibitem{shiu2000fading}
D.-S. Shiu, G.~J. Foschini, M.~J. Gans, and J.~M. Kahn, ``Fading correlation
  and its effect on the capacity of multielement antenna systems,'' \emph{IEEE
  Trans.\ Communications}, vol.~48, no.~3, pp. 502--513, 2000.

\bibitem{lee2001algorithms}
D.~D. Lee and H.~S. Seung, ``Algorithms for non-negative matrix
  factorization,'' in \emph{Proc.\ NIPS}, Dec. 2001, pp. 556--562.

\bibitem{nocedal2006numerical}
J.~Nocedal and S.~J. Wright, ``Numerical optimization,'' \emph{Numerical
  optimization}, pp. 497--528, 2006.

\bibitem{3GPP36.814}
3GPP, ``{Further advancements for {E-UTRA} physical layer aspects},'' TR 36.814
  (release 9), 2010.

\bibitem{ITU-M.2134}
ITU, ``M.2134: {R}equirements related to technical performance for
  {IMT-A}dvanced radio interfaces,'' Technical Report, 2009.

\bibitem{Madhow:ADC}
H.~Zhang, S.~Venkateswaran, and U.~Madhow, ``Analog multitone with interference
  suppression: Relieving the {ADC} bottleneck for wideband 60 {GHz} systems,''
  in \emph{Proc.\ IEEE Globecom}, Nov. 2012.

\bibitem{Madhow:largeArray}
D.~Ramasamy, S.~Venkateswaran, and U.~Madhow, ``Compressive tracking with
  1000-element arrays: a framework for multi-gbps mm wave cellular downlinks,''
  in \emph{Proc. 50th Ann. Allerton Conf. on Commun., Control and Comp.},
  Monticello, IL, Sep. 2012.

\bibitem{Nesterov:07}
Y.~E. Nesterov, ``Gradient methods for minimizing composite objective
  function,'' \emph{CORE Report}, 2007.

\bibitem{Rangan:11-ISIT}
S.~Rangan, ``Generalized approximate message passing for estimation with random
  linear mixing,'' in \emph{Proc. IEEE Int. Symp. Inform. Theory}, Saint
  Petersburg, Russia, Jul.--Aug. 2011, pp. 2174--2178.

\end{thebibliography}

\end{document}